# Modified K-means with Cluster Assignment - Application to COVID-19 Data

Shreyash Rawat[1], V. Vijayarajan[1], V. B. Surya Prasath[2,3,4,5]*

[1]School of Computer Science and Engineering (SCOPE), Vellore Institute of Technology, Vellore, India
shreyash.rawat@gmail.com, vijayarajan.v@vit.ac.in
[2]Division of Biomedical Informatics, Cincinnati Children's Hospital Medical Center, Cincinnati, OH 45229 USA
[3]Department of Pediatrics, University of Cincinnati, OH USA
[4]Department of Biomedical Informatics, College of Medicine, University of Cincinnati, OH 45267 USA
[5]Department of Computer Science, University of Cincinnati, OH 45221 USA
*prasatsa@uc.edu

**Abstract**

Text extraction is a highly subjective problem which depends on the dataset that one is working on and the kind of summarization details that needs to be extracted out. All the steps ranging from preprocessing of the data, to the choice of an optimal model for predictions, depends on the problem and the corpus at hand. In this paper, we describe a text extraction model where the aim is to extract word specified information relating to the semantics such that we can get all related and meaningful information about that word in a succinct format. This model can obtain meaningful results and can augment ubiquitous search model or a normal clustering or topic modelling algorithms. By utilizing new technique called two cluster assignment technique with K-means model, we improved the ontology of the retrieved text. We further apply the vector average damping technique for flexible movement of clusters. Our experimental results on a recent corpus of Covid-19 shows that we obtain good results based on main keywords.

## 1. Introduction

We live in an information rich world where we are inundated with a large number of ambient resources like text data and we typically fail to leverage all of it. There is a gargantuan amount of information present on the world wide web whether it be the books or articles. There is a huge amount of gap between the skills we possess to make use of the information and the amount of information available. Large amounts of data tacitly lead to large amounts of information that may or may not be useful for real world insights, hence we need tools and techniques to refine the data available to us so that we can exactly get what we want. Scientific research articles in journals and books are one major area where a lot of potential information exists. However, due to the large number of papers on each topic it is almost impossible for one person to go through all of them and get the information that she was looking for about a specific technique or a new invention. Effective text extraction is one of the solutions to this important problem. There has been a tremendous amount of research on text extraction and we have surely come a long way from the simple keyword search algorithm to the various complex algorithms for effective topic modelling. The one area that has been left scarce is word weighted clustering where you get the power of clustering algorithms along with the topics that you want to extract information about [17]. Normal clustering such as the ones based on K-means or topic modelling algorithms take each input token (derived from sentence) as an equal entry and clusters the points closest in the region [7]. This method does give you the most frequently occurring topics but might well miss out on topics that are very important but less frequently occurring. In this work, we consider the classical K-means clustering as the basis for text mining from scientific literature.

With the recent coronavirus disease (COVID-19), there is a deluge of scientific articles that are being released and effective summarization and clustering are important in distilling the salient information [5]. The Kaggle COVID-19 open research dataset (CORD) [16], indexed more than 63000 scholarly articles including over 51000 with full text containing information about the COVID-19, SARS-CoV-2 and related coronaviruses. There is an immense need for effective clustering of these scientific articles to extract important details for the medical and research community in tackling this ongoing pandemic. Using clustering models within the text mining literature we propose to create an efficient model wherein information about COVID-19 related topics can be extracted by giving importance to that word in determining the cluster centroids in the algorithm. Leveraging the K-means clustering model along with our newly introduced two cluster assignment technique here, we devise a keywords-based summarization of COVID-19 data. We further utilize the classical vector averaging damping factor [2] for free movement of the cluster centroids we obtain accurate results for not just the frequent topics but also for the important ones that the user wishes to search for. Our experimental results indicate that we obtain good results based on the modified clustering approach considered here. Further, relevant keywords-based text mining on the recent COVID-19 Open Research Dataset (CORD-19) dataset indicate that the two-cluster assignment along with K-means provided succinct and salient summarized information from the large amount text.

We organized the rest of the paper as follows. Section 2 introduces our modified K-means model for clustering the new COVID-19 dataset based on important keywords. Section 3 provides experimental results and analysis, and Section 3 concludes the work.

## 2. Methodology

2.1 Dataset

The dataset we used here consists of 2GB text data on coronavirus (COVID-19, SARS-CoV-2) COVID-19 Open Research Dataset (CORD-19) [6], which has over 59000 scholarly articles and includes over 41000 with full text availability. This is an ideal data to test our algorithm as the dataset is fresh due to the velocity of the amount of research done in a very short period of time. We have taken 100/200 research papers from 3 different corpus namely Biorxiv_medxriv, comm_use_subset and pmc_custom_license. We obtained similar results with batch sizes 400 and above. The subsets allow us to easily visualize the data points without crowding and helps us visualize the clustering process from a fundamental view. Hence the results can be used by the researchers to extract information rather than the prosaic keyword search algorithms [16]. This dataset contains information not only about coronavirus but also every related article relating to the coronaviruses and therapeutics. Due to the rapid research perpetuating, these techniques will be in demand and can be required to extract meaningful information as quickly as possible.

2.2 Preprocessing

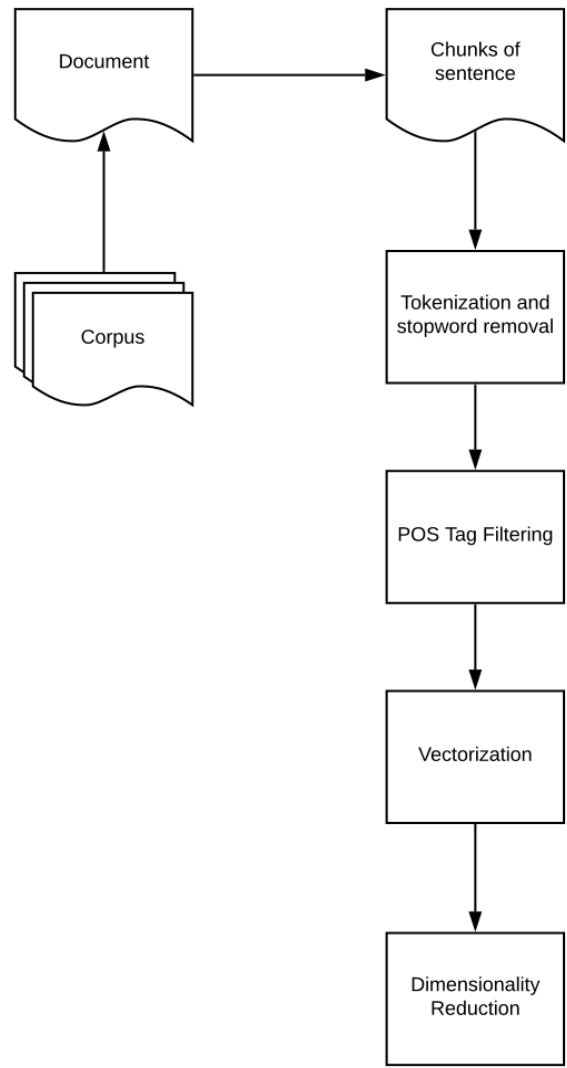

Fig.1. Flowchart of the pre-processing steps utilized in our clustering approach for Covid-19 scholarly articles.

Preprocessing steps are important as they set up the data appropriately for stages of text mining and information extraction. The cleaner the data the better will be the results predicted by the model. This is also one of the most obscure steps among all as there is no specified steps or framework to carry out preprocessing. It largely depends on the context of the problem and the dataset that is being analyzed for insights [8]. The overall steps we have utilized here are shown in Figure 1. In our work, the preprocessing starts by chunking the research paper into a group of two to three sentences. Taking the assumptions that two sentences will probably be stating similar topics, this technique has two benefits, the first being the reduction of time complexity as instead of assigning each new sentence we will assign chunks that will probably give similar outputs. Effective clusters due to absence of redundant data points is the second benefit that this method possesses. Then these chunks have to be arranged in a structured format to feed them into machine learning models. This is done by first tokenizing each chunk and removing the stop words and punctuations from the sentences. In research articles we can find a lot of words that do not require attention when it comes to text extraction like numbers, websites, citations, figures and graphs. Since we are dealing with scholarly articles there were a lot of specific kinds of words that were to be kept and many that were removed. Regular expressions were used for this purpose wherein certain patterns were compiled, and the tokens were then matched with these expressions [9].

Apart from removing the regular stop words, there were still many words like articles or numbers that have no use to cluster sentences but will unnecessarily increase the dimensions of the vector embeddings of these sentences. Hence to dispel such words we used a POS eliminator wherein we remove words with certain parts of speech to generate accurate vectors according to the main topics. This played an important part in removing the redundant pronouns and articles. After getting the tokens cleaned, we now have to create vectors for these tokens so that they can be fed into the predictive models. There are various kinds of approaches present to facilitate the token to vector transformation like the Count Vectorizer, tf-idf and the word to vec. We have incorporated the tf-idf approach [10] for two important reasons:
1. It reduces the importance of the frequency for a given word. If we had used CountVectorizer, it would have given huge importance for frequently occurring words which is exactly what we do not want. Hence, it reduces the weight factor of frequency while determining the vector values.
2. It tf-idf is a normalized model and obtains the values within a range so that the model does not categorize the important words with a high weight vector as outliers.

2.3 Dimensionality Reduction and Batching

While generally converting sentences into tf-idf vectors the number of columns represent the vocabulary of the training set which is usually very large as it takes into account all the cleaned tokens of all the sentences. This is not favorable to the clustering algorithms as it increases the time required to calculate the distance between each point and the cluster centroid [11]. Another drawback of a large number of dimensions is the presence of redundant or similar words that are not lemmatized to the same root word. It takes words like these as separate columns which eventually mean the same thing and unnecessarily increases complications. Dimensionality reduction is also required to visualize the data as it is elusive to make sense of data by just looking at their vectors. The data visualization for the keywords "Vaccine" and "Transmission" are shown in Figure 2.

There are plenty of dimensionality reduction algorithms present for this purpose like principal component analysis (PCA), linear discriminant analysis (LDA) and auto encoders [12]. While autoencoders are generally suited for image and audio data, LDA explores the relationship between dependent and independent variables. In other words, it finds the direction that maximizes the difference between two classes which is ideal for prediction purposes. But what we want here is the distinction between independent variables for clustering purposes and PCA does exactly that. PCA finds the direction which maximizes the variance in the data thus discovering the relationship between independent variables.

It is cumbersome to store and process all the available data in the memory as it exhausts the RAM and no further processing is possible on data. To overcome this obstacle, we have used a batching technique wherein the entire dataset is stored in the memory rather than the RAM and we access the memory in batches, process it and then store it back to the memory. This prevents the RAM memory from being exhausted although it may increase the processing time to some extent. We have utilized batch sizes of 50 full text scholarly articles which will be processed at one batch. Another major advantage of our modified K Means algorithm is the time complexity. The additional processing steps coalesced into the algorithm does not increase the time complexity by polynomial factors. The time complexity of our algorithm turns out to be $O(n^2 + n)$ which approximates to the time complexity of the normal K-Means which is $O(n^2)$.

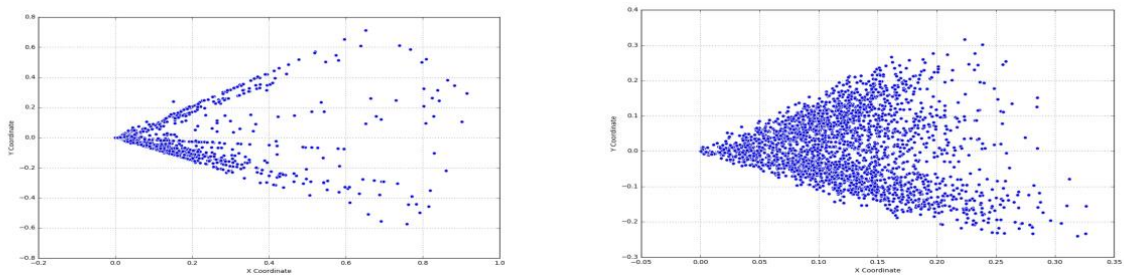

Fig 2. Distribution of COVID-19 scholarly articles data with search query keywords as "Vaccine" and "Transmission".

2.4 Clustering

For forming effective clusters from the normalized tf-idf vectors as inputs we used a variation of K-means algorithm. The standard K-means algorithm [13] assigns K number of random points as clusters having equal weights and then finds the distance of each point from the clusters and then the new cluster centroids are derived by taking the average of all the points belonging to that particular cluster. This process goes on until the average of the cluster points is the same as the previous clusters. This method works well for most common cases and gives you the most commonly occurring topics. Hence, if we want a cluster with a specific topic, we need to keep increasing the number of clusters until we get that cluster.

To overcome this obstacle we have come up with a modified K-means model that has three amendments to improve performance. The major one being giving explicit weights to each tf-idf vector [15]. The weights that will be assigned to each vector will depend on the search query or the word that one is interested in. The chunks where this word is syntactically and semantically significant gets a higher score compared to others. The rest of the chunks that are not related to the search query get a default score of 0 which causes quite a few problems. The cloistered chunks are given a weight of 0.01 in our case and may vary depending upon the nonzero weights of the other chunks. We need to ensure that the weighted chunks that are related to the search term should discern from the others.

If we give all the cloistered chunks a weight of zero, then the clusters obtained will all be resolutely focused on the search query which will lead to redundant and repeating clusters containing the same information. Hence the chunk with a higher weight pulls the centroid of that cluster more strongly than the rest which leads to clusters biased towards these vectors, see Figure 3 for the visualization. If we assign the unrelated chunks with a weight of 0, then they will not contribute anything towards the calculation of new centroids for that epoch. This will lead to all the clusters being crowded near the search term and will lead to inaccurate results. To overcome this impediment the unrelated chunks were assigned a weight of 0.01 so that they have a minimal say during the centroid calculation but do not influence the centroids to a large extent. The other amendment to the K-means we proposed includes a novel technique called "two cluster assignment". The motivation behind this technique was the inclusion of chunks under more than one topic or cluster. Since we are using chunks that greatly reduces the computation power and time complexity, there is a possibility that a group of sentences belong to multiple clusters. Therefore, while assigning a chunk to a cluster, if the distance between the chunk and the two centroids is less than a threshold then we assign that chunk to both the clusters. This leads to more accurately extracted text for both the topics. Another change that we have incorporated is the usage of centroid damping while calculating the mean of cluster centroids. This technique is similar to the average damping factor [1] but instead of assigning a weight of zero to the current centroid, it gives it a weight that has been assigned to the cloistered chunks which is 0.01 in our case. This neither devalues the current cluster centroid nor exalts it, making it ideal for the calculation of the new cluster centroid. Algorithm 1 describes the steps of our two-cluster assignment-based K-means model.

Iteration-wise clustering for the words "Vaccine", "Transmission", "Incubation", and "Surveillance" are shown in Figure 4 to 7 respectively. The black points shown in these images represent those points which were assigned to two clusters simultaneously in that iteration. After the centroids stabilize, the highlighted common points were usually those which fit well in both cluster topics hence making the cluster content more accurate than before. After obtaining the cluster centroids, we now need to assign each chunk to a cluster so that we can retrieve the full text of the required cluster. The results obtained were quite accurate stating results about the search query in great depth. We note that clusters about the same topic have been well separated explaining different things about the same topic.

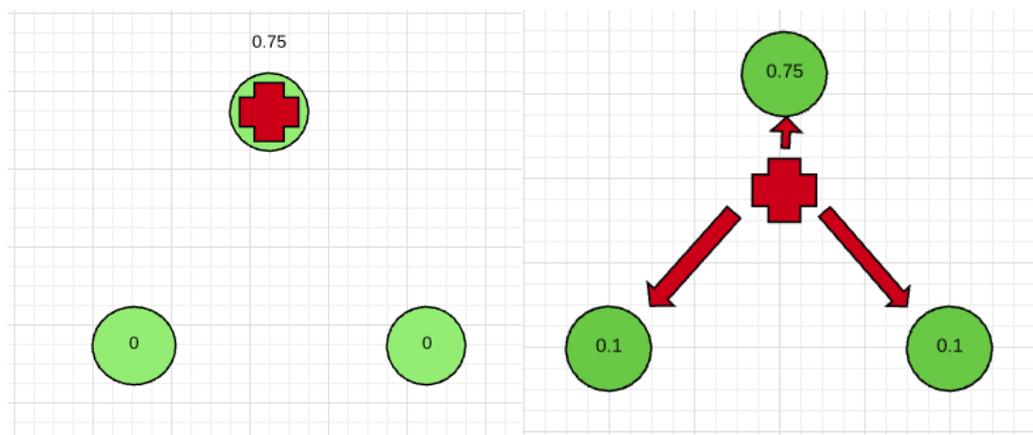

Fig 3. Weights assignment for two cluster assignment-based K-means model. We assign two clusters instead of one.

Algorithm 1

1. $t = 0$
2. Set $\{c_1^t, c_2^t, \ldots, c_k^t\}$ to be distinct points from input $\{x_1, x_2, \ldots, x_n\}$

3. Repeat
    t = t + 1
    d = []
    for i = 1….n do
        for j = 1…..k do
            d.append (|| cj^t, xi ||, cj^t)
        x = sort(d[0])
        j* = x[0][1]
        cj* = cj* U {xi}
        If (x[1][0] - x[0][0] < 0.01) #threshold
            i* = x[1][1]
            ci* = ci* U {xi}
4. #Cluster assignment
    Foreach i = 1….k do
        ri^t = (1 / |Ci| ) * sum( xj * wj ) for every xj in Ci
        Until sum from 1 ...k  ( || ri^t - ri^(t-1)) < epsilon
    end

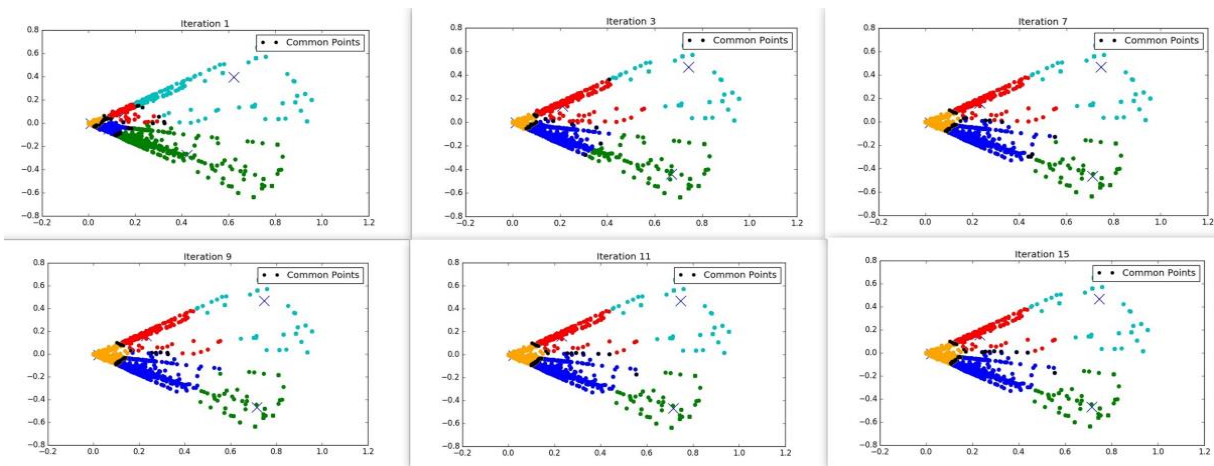

Fig 4. Iteration-wise clustering for the word "Vaccine". We show for iterations 1, 3, 7, 9, 11, 15. The black points represent the common cluster points through the iterations.

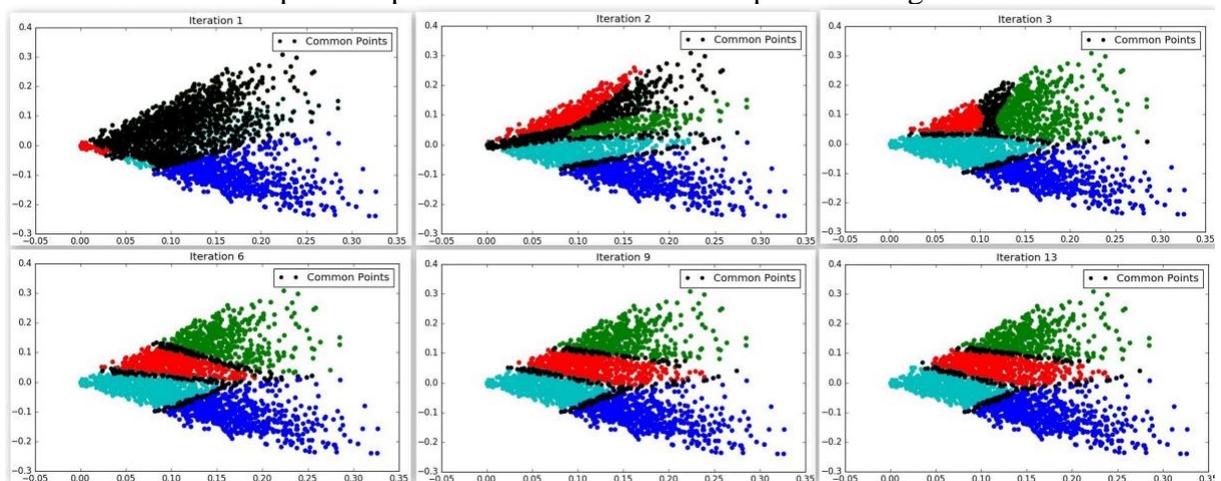

Fig 5. Iteration-wise clustering for the word "Transmission". We show for iterations 1, 2, 3, 6, 9, and 13. The black points represent the common cluster points through the iterations.

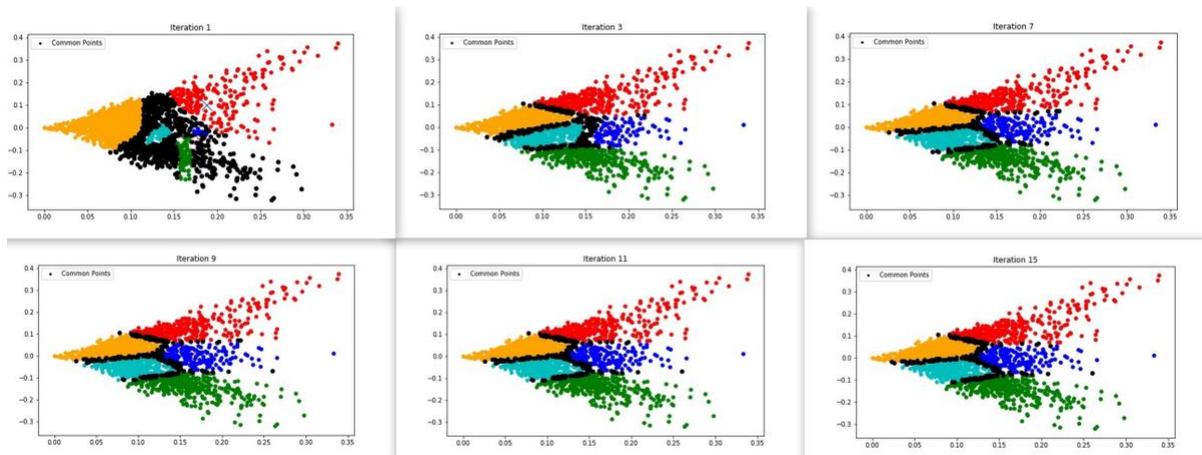

Figure 6. Search query - "Incubation". We show for iterations 1, 3, 7, 9, 11 and 15. The black points represent the common cluster points through the iterations.

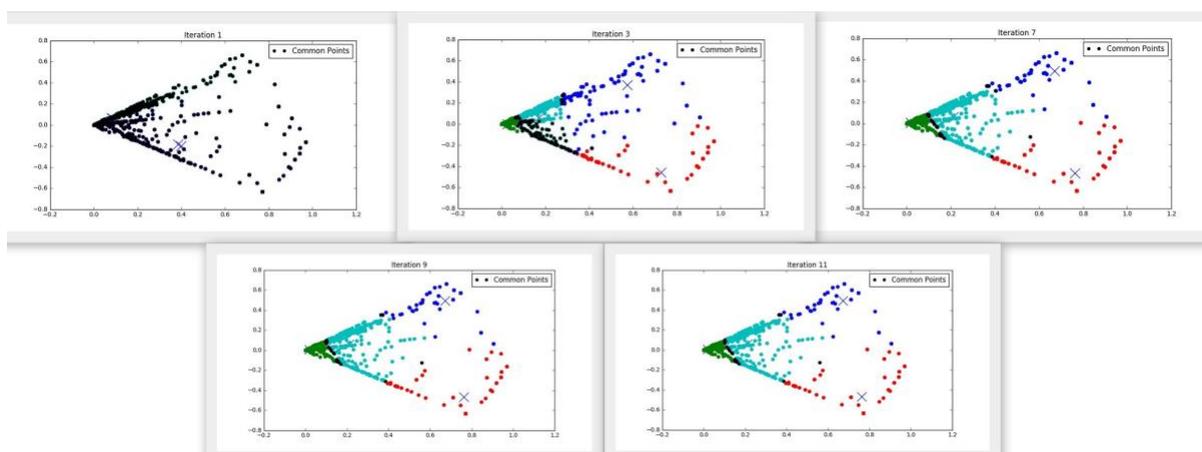

Figure 7. Search query - "Surveillance". We show for iterations 1, 3, 7, 9, and 11. The black points represent the common cluster points through the iterations.

2.5 Fine Tuning

Selecting the ideal value of K for the algorithm has always been a tedious task. There are few methods like the elbow method [14] that plot the distortion score against the number of clusters like those shown in Figure 8. The ideal number of clusters turns out to be the one from where the graph starts to flatten out. The distortion score is a metric which is calculated by how far each point is from its assigned cluster. While following the "two cluster assignment" technique there can be one point that is a part of two clusters. In that case the point will be counted in both the clusters separately to determine the distortion factor.

Table 1. Results obtained for word "Vaccine"

| Corpus | Total no. of para | # of related para obtained by search algorithm | # of related para obtained by normal K-Means | # of related para obtained by modified K-means |
|---|---|---|---|---|
| Biorxiv_medxriv | 3785 | 1758 | 1150 | 1156 |
| comm_use_subset | 2115 | 948 | 474 | 479 |
| pmc_custom_license | 2012 | 1023 | 452 | 461 |

Table 2. Results obtained for word "Transmission"

| Corpus | Total no. of para | # of related para obtained by search algorithm | # of related para obtained by normal K-Means | # of related para obtained by modified K-means |
|---|---|---|---|---|
| Biorxiv_medxriv | 3785 | 1615 | 898 | 915 |
| comm_use_subset | 2115 | 1024 | 732 | 733 |
| pmc_custom_license | 2012 | 893 | 391 | 396 |

Table 3. Results obtained for word "Incubation"

| Corpus | Total no. of para | # of related para obtained by search algorithm | # of related para obtained by normal K-Means | # of related para obtained by modified K-means |
|---|---|---|---|---|
| Biorxiv_medxriv | 3785 | 899 | 710 | 717 |
| comm_use_subset | 2115 | 561 | 337 | 370 |
| pmc_custom_license | 2012 | 566 | 289 | 311 |

Table 4. Results obtained for word "Surveillance"

| Corpus | Total no. of para | # of related para obtained by search algorithm | # of related para obtained by normal K-Means | # of related para obtained by modified K-means |
|---|---|---|---|---|
| Biorxiv_medxriv | 3785 | 153 | 79 | 83 |
| comm_use_subset | 2115 | 264 | 115 | 127 |
| pmc_custom_license | 2012 | 248 | 164 | 164 |

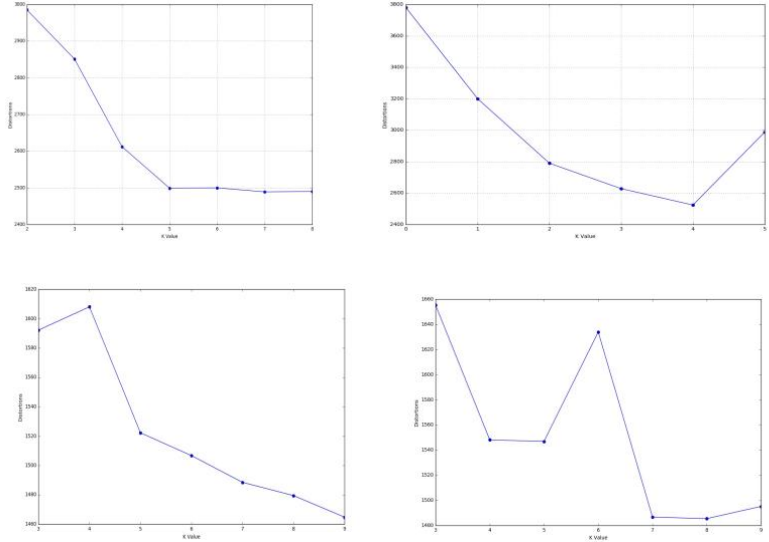

Fig 8. Trend of distortions values as the K-values increase, distortions versus K values for keywords: (a) "vaccine", (b) "transmission", (c) "incubation", (d) "surveillance".

The number of dimensions the tf-idf vector should be reduced to is again an empirical decision where we can only get the answer by implementing with different values and checking the distortion values of the clusters. The max number of iterations of the algorithm is another factor to be looked upon. It is usually set to a higher value to be safe about the right clusters being formed. Partial clustering [2] is one powerful method which can be a substitute to random seeding of the centroids. In this method, K-means is performed on a subset of data and then those obtained clusters act as the initial clusters for the entire data. This might be more effective than random seeding but there will be a huge difference in the clustering time as you perform the algorithm 1.5 times instead of 1.

### 3. Experimental Results and Discussion

After obtaining the clusters, we need to check if the results obtained corroborate with the required information. In our experiments, we have chosen some example weighted words, such as "Vaccine" and "Transmission", "Incubation", and "Surveillance" as we wanted some useful information about them. About vaccines, the first task consists of finding all possible things that we know about the vaccines for coronavirus and the second being the methods or procedures that can be applied for making the future vaccine that can cure this disease. Similarly, for the transmission we were looking for information about its transmission rate and means of transmission. We can judge our accuracy of the clusters by checking the ten most common words occurring in each cluster which will give us a good idea as to what kind of sentences the clusters accommodate.

We got good results wherein two clusters had the most important word as vaccine. One which gave insights about the vaccine and what was currently being used and the second cluster consisted of the potential combinations that could form the vaccine for the coronavirus. The only way to judge the accuracy of the clustering algorithm in this kind of problem is by trying out different K values until you get the required clusters. The threshold value to be selected in the two-cluster assignment technique is another empirical decision to be made. We tried the model for various threshold values according to the distances obtained for each point from each cluster centroid. The most vigilantly separated results were obtained for a threshold value of

0.01 where the points were neatly clustered, and the common points were important to both the clusters.

The number of related paragraphs obtained for the sample queries have been portrayed in Table 1 to 4. There is a conspicuous difference between the modified K-means and the search algorithm, where the obtained information in case of K-means is concise and well related to the cluster hence facilitating accurate text extraction by winnowing out the junk which does contain that word but is not related to the context. The Figure 9 shows the explanation obtained for the invention of the new vaccine that can be proposed. As evident from the Figure 9 and Figure 10, there are numerous chunks that are close to each other and that cluster constitutes common information regarding vaccines. As the points deter, they form two to three other clusters which contain specific information or topics on vaccines like scope for invention of new vaccines or current knowledge of vaccines to fight different kinds of viruses. This is clearly evident as the results obtained in Figure 9 states about the possible new structure for invention of vaccines and Figure 10 gives us information about the vaccines and technology currently in practice. Figure 11 to 13 shows the content obtained by taking the words "Transmission", "Incubations", and "Surveillance". Figure 11 shows results for "Transmission" as the search query and we note that it provides relevant information about the transmission information both experimental and academic about the novel coronavirus.

> The best predicted vaccine from the previous steps, was reverse transcribed to a possible DNA sequence which is supposed to express the vaccine protein in a target organism. The cellular machinery of that particular organism could use the codons of the newly adapted DNA sequence efficiently for producing the desired vaccine. Codon adaptation is a necessary step of vaccine design because this step provides the effective prediction of the DNA sequence of a vaccine
>
> construct. An amino acid can be encoded by different codons in different organisms, which is known as codon bias. Codon adaptation predicts the best codon for a specific amino acid that . CC-BY 4.0 International license author/funder. It is made available under a The copyright holder for this preprint (which was not peer-reviewed) is the . https://doi.org/10.1101/2020.02.05.935072 doi: bioRxiv preprint should work effectively and efficiently in a specific organism. The best predicted vaccine was used for codon adaptation by the Java Codon Adaptation Tool or JCat server (http://www.jcat.de/). The server ensures the maximal expression of protein in a target organism. Eukaryotic E. coli strain K12 was selected at the JCat server and rho-independent transcription terminators, prokaryotic ribosome binding sites and SgrA1 and SphI cleavage sites of restriction enzymes, were avoided.
>
> The MHC class-I and MHC class-II epitopes, determined for potential vaccine construction. The IEDB (https://www.iedb.org/) server generates a good number of epitopes. However, based on the antigenicity scores, ten epitopes were selected from the top twenty epitopes because the epitopes generated almost similar AS and percentile scores. Later, the epitopes with high antigenicity, nonallergenicity and non-toxicity were selected for vaccine construction. The B-cell epitopes were also selected based on their antigenicity, non-allergenicity and length (the sequences with more than 10 amino acids). Table 04 and Table 05 list the potential T-cell epitopes of nucleocapsid phosphoprotein and Table 06 and Table 07 list the potential T-cell epitopes of surface glycoprotein. Table 08 lists the predicted B-cell epitopes of the two proteins and Table 09 lists the epitopes that followed the mentioned criteria and were selected for further analysis and vaccine construction.
>
> The current study was designed to construct possible vaccines against the Wuhan Novel Coronavirus 2019 (SARS-CoV-2), which is the cause of the recent outbreak of the deadly viral disease, COVID-19 in China. The pneumonia has already caused the death of several thousands of people worldwide. For this reason, possible vaccines were predicted in this study to fight against this lethal virus. To carry out the vaccine construction, four candidate proteins of the virus were identified and selected from the NCBI database. Only highly antigenic sequences were selected for further analysis since the highly antigenic proteins can induce better immunogenic response [185, 201]. Because the nucleocapsid phosphoprotein and surface glycoprotein were found to be antigenic, they were taken into consideration for vaccine construction.

Fig 9. Gives information about the potential new vaccines

Molecular docking analysis is one of the essential steps in reverse vaccinology to design vaccines.

The best predicted vaccine from the previous steps, was reverse transcribed to a possible DNA sequence which is supposed to express the vaccine protein in a target organism. The cellular machinery of that particular organism could use the codons of the newly adapted DNA sequence efficiently for producing the desired vaccine. Codon adaptation is a necessary step of vaccine design because this step provides the effective prediction of the DNA sequence of a vaccine

The protein-protein docking study was carried out to find out the best constructed COVID-19

vaccine. The vaccine construct with the best result in the molecular docking, was considered as the best vaccine construct. According to docking results, it was found that CV-1 was the best constructed vaccine. CV-1 showed the best and lowest scores in the docking as well as in the MM-GBSA study. However, CV-2 showed the best binding affinity (ΔG scores) with DRB3*0202 (- showed the best results in the protein-protein docking study, it was considered as the best vaccine construct among the three constructed vaccines (Figure 09 & Table 15 ). Later, the molecular dynamics simulation study and in silico codon adaptation studies were conducted only on the CV-1 vaccine. The copyright holder for this preprint (which was not peer-reviewed) is the . https://doi.org/10.1101/2020.02.05.935072 doi: bioRxiv preprint

For this reason, the production of the CV-1 vaccine should be carried out efficiently [183] [184] .

After the successful docking study, the vaccine construction was performed. The linkers were used to connect the T-cell and B-cell epitopes among themselves and also with the adjuvant sequences as well as the PADRE sequence. The vaccines, with three different adjuvants, were constructed and designated as: CV-1, CV-2 and CV-3. Since all the three vaccines were found to be antigenic, they should be able to induce good immune response. Moreover, all of them were possibly nonallergenic, they should not be able to cause any allergic reaction within the body as per in silico The copyright holder for this preprint (which was not peer-reviewed) is the . https://doi.org/10.1101/2020.02.05.935072 doi: bioRxiv preprint antigen. Moreover, the conventional approach of vaccine development have raised many safety concerns in many pre-clinical and clinical trials. The subunit vaccines like the vaccines predicted in the study could overcome such difficulties [196] - [200] . Finally, this study recommends CV-1 as the best vaccine to be an effective worldwide treatment based on the strategies employed in the study to be triggered against SARS-CoV-2 infection. However, further in vivo and in vitro experiments are suggested to strengthen the findings of this study.

Fig 10. Gives information on current knowledge of vaccines

A susceptible-exposed-infectious-quarantine model was used for transmission analysis and prediction of epidemiological spread,

is the (which was not peer-reviewed) The copyright holder for this preprint . https: //doi.org/10.1101 //doi.org /10. /2020 where S, E, I, Q and N were the number of susceptible (S), exposed (E), infectious (I), quarantined (Q) and total population. The population in Wuhan (N) is 11,081,000. Assuming that a patient was quarantined immediately after the diagnosis was confirmed, Q was equal to the confirmed number of cases. β is the daily transmission rate, defined as the expected number of infections caused by one infectious person per day. Once a susceptible (S) person becomes infected, the status is changed to exposed for an incubation period, 1/α. Theoretically, the patient is not infectious during the incubation period. After incubation, the patient experiences disease onset and becomes infectious (I). The time interval between disease onset to quarantine (Q) is the infectious time, 1/γ.

Most of infected cases in December 2019 were linked to the Huanan Seafood Market, which retails seafood and wild animals. It is believed that an enclosed and crowded environment is favorable for coronavirus transmission, however the epidemiological data is lacking. To test the hypothesis, the recent Diamond Princess cruise epidemic was used as a comparable case study. A clear disease outbreak was reported on this cruise. A passenger, who visited China on January 10 th , was on board from January 20 th and January 25 th before being confirmed with a SARS-CoV-2 infection. All people on board have been quarantined at sea since February 5 th and 621 out of 3711 people were confirmed positive for SARS-CoV-2. As some of the patients were known to be infected after the quarantine, possibly due to central air conditioning and family infection, only the confirmed cases (n=135) by February 10 th (five days incubation plus one day diagnosis) were used for a conservative estimation of transmission. Using a SEIQ model while Q=0, the daily transmission rate, β, was found to be 1.04 (95% CI 0.69-1.87) ( Figure 1B ), which is twice as much as the transmission rate in open cities. Using the infection period of 12.5 days, the effective reproductive number on the cruise, Rc, was 13.0 (95% CI 8.63-23.375).

is the (which was not peer-reviewed) The copyright holder for this preprint . https://doi.org/10. 1101 To characterize the zoonotic infection of COVID-19, a SEIQ model was constructed to analyze for the epidemic from the Hunan Seafood Market in December 2019. The first SARS-CoV-2 onset was found on December 1 st 2019 (Huang et al., 2020) , and by the 1 st of January, 41 patients were confirmed and quarantined. Using the 5.8-day interval between disease onset and clinical visit, 136 cases diagnosed in the first 6 days of January were included, bringing the total number of infections to 177. Given β0=0.44, the development of the epidemic was simulated by initializing with a range of numbers (from 1 to 5) of infections on December 1 st using the SEIQ model. It showed that at most one infected patient can be allowed in the model, which induced a mean of 174 infections (95% CI 161-187) by January 1 st . Instead of one, if two unrelated people were infected by December 1 st , the 95% CI would be (291, 318) on January 1 st , significantly larger than the expected 177 cases. Thus, the transmission would be so minimal that it would not substantially contribute to the final number of infections, even if there existed a second zoonotic source. In the same sense, the results did not support a continuous zoonotic source within the Huanan Seafood Market, which would have resulted in a higher overall number of infections in the later trajectory.

Fig 11. Results for cluster with most important word as "Transmission"

By February 8 th there were 37,198 confirmed cases nationally, with 27,100 of these cases in Hubei (72.9% of all cases), and 14,982 in Wuhan (40.2% of all cases). All models applied to these data estimated the basic reproduction number effectively. Basic reproduction numbers for all fives methods for the entire time period, the pre-closure period and the post-closure period, are shown in Table 1 . The best-fitting method in the entire period was the method based on time-dependent reproduction numbers, while the pre-closure and post-closure period were best fitted by the exponential growth model.

The 2019-nCoV epidemic has spread across China and 24 other countries 1-3 as of February 8, 2020 . The mass quarantine measure implemented in the epidemic's epicenter, Wuhan, Hubei Province, China, on January 23, 2020, has stopped the human migration between Wuhan and other places. However, according to the local authorities, over five million people had already traveled from Wuhan to other parts of China in the weeks leading to January 23, 2020, with the majority (79.96%) to the other cities within Hubei.

The Figure presents the estimated number of infected individuals, as well as the healthcareresource-adjusted vulnerability to the 2019-nCoV outbreak across mainland China, as of January 29, 2020. The epidemiological parameters may have changed after January 23, 2020, when the lockdown measure was implemented. Thus, we only use the data by January 23, 2020 and produce the estimation till January 29, 2020 to accommodate the six-day incubation period identified by Wu et al. 5 .

is the (which was not peer-reviewed) The copyright holder for this preprint . https://doi.org/10.1101/2020.02.14.20022913 doi: medRxiv preprint 1 0 ~is set to take effect on four time: January 9, January 16, January 23, January 30. The x-axis is the date and the y-axis is the number of patients.

On 12 January 2020, the World Health Organization (WHO) declared the novel coronavirus which 54 caused unknown pneumonia cases in Wuhan, Hubei Province, China since December 2019 as "2019-55 nCoV", which was renamed by the International Committee on Taxonomy of Viruses as "SARS-CoV-2" 56 on 11 February 2020. In parallel, the WHO formally named the disease caused by SARS-CoV-2 as 57 "COVID-19", short for Coronavirus Disease 2019. Back in late December 2019, a cluster of 27 58 pneumonia cases associated with SARS-CoV-2 with a common link to the Huanan Seafood Wholesale 59 Market were reported [1] , and the first death case attributable to SARS-CoV-2 occurred on 9 January 60 2020. Soon after the first global incidence was confirmed in Thailand on 12 January 2020, new cases 61 were reported in different countries and were mostly associated with Wuhan travel history or residency. To control this COVID-19 epidemic, much effort has been paid to identifying the etiological agent, 77 epidemiological parameters such as incubation period [4] , disease transmissibility [4, 5] , clinical 78 . CC-BY-NC-ND 4.0 International license It is made available under a author/funder, who has granted medRxiv a license to display the preprint in perpetuity.

Figure 12. Results obtained for the keyword "Incubation"

Various definitions or criteria of an EO have been used in different outbreaks depending on the type of disease and the institutions providing technical guidance or in charge of controlling the outbreak. We retrieved examples of EO criteria from various pathogens and institutions from the literature ( Table 1) . Most of these criteria are obtained from the World Health Organization (WHO) guidelines or decided by a government's Ministry of Health or public health authorities. The most commonly used EO criterion is a period of twice the longest incubation period without observing any new cases since the last possible transmission event (4) (5) (6) (7) (8) (9) (10) (11) (12) . In an Ebola Virus Disease (EVD) outbreak context, the EO of EVD is declared after 42 days (twice the longest incubation period) has passed since the outcome of the last case. There are several possible scenarios of the outcome of the last case (8) . If the last case is a laboratory-confirmed case, the outcome can be the second PCR negative test of blood samples or a safe burial if the person died. Additionally, if the last case is a contact of a confirmed case, the outcome is a safe burial after the person's death.

Secondly, the use of the maximum incubation period is challenging with typically limited sample sizes available for estimation, and it does not produce any probabilistic risk assessment (18) . Lastly, it is important to consider missing cases due to imperfect surveillance during the outbreak and asymptomatic cases that potentially act as invisible transmission sources during the outbreak and could prolong the time to declare the EO (18, 19) .

However, these approaches only address some of the aforementioned issues and have other limitations which we discuss later. Hence, further development of quantitative techniques to define the EO is urgently needed. In this study, we developed a simulation-based method to calculate the probability of cases arising in the future after the outcome of the last detected case. We account for factors that influence the estimated probability such as the underlying reproduction number, the reporting rate and the delay between the onset and the outcome of the last detected case, which for simplicity we will refer to as 'the onset-to-outcome delay period'. We tested our method on several EVD outbreak scenarios. Finally, we used the simulation results to propose a new quantitative criterion to define the EO for EVD.

We developed a quantitative framework to determine the timing of an EO declaration, which divides the outbreak into three phases: the outbreak phase; the onset-to-outcome delay phase; and the EO declaration phase (Figure 1 ). The outbreak phase encompasses the outbreak trajectory which includes all detected cases up until the time that the EO analysis is about to be conducted. The onset-tooutcome delay phase is the period between the onset of the last detected case to the outcome (recovery or death) of that case. In this phase, there is a risk of undetected cases sustaining transmission beyond the last detected case due to underreporting.

Accounting for these potential 'invisible' sources of transmission is important to determine the EO with confidence, irrespective of potential underreporting. The EO declaration phase starts the day after the outcome of the last case. In this phase, the probability of cases arising in the future is calculated for each day until a predefined probability threshold, here fixed at 5%, is reached.

Fig 13. Results obtained for search term as "Surveillance"

## 4. Conclusion

We presented a word weighted information extraction algorithm that retrieves information not only syntactically but also semantically giving much more accurate results than the prosaic search algorithm. We introduced quite a few modifications in the existing K-Means algorithm which in turn proved out to be quite effective. The introduction of chunking for large datasets proved out to be a tenable step which not only helped in reduction of time complexity but also made predictions much accurate simultaneously reducing computing power. The two-cluster assignment technique made sure that group of sentences is present in all the clusters that it should be in rather than the one it is closest to, making our approach efficient. The tweaked word weight vector that assigns the non-related vectors a partial weight rather than zero provided good results by not undermining the non-related vectors completely. The performance of our proposed clustering method can further be improved. We can use other effective seeding methods that reduces the clustering time. Another proposed information retrieval technique can be retrieval of information using semantic similarity. If we can represent a sentence into a vector that shows its semantic properties, then that can be used to get sentences similar to a search query. The only difference will be the input entered will be a sentence instead of a word.